\documentclass[final]{siamart1116}
\usepackage{amsfonts}

\def \set#1{\ifmmode \{\,#1\,\}\else $\{\,#1\,\}$\fi}
\def \q#1 {\bigskip \noindent #1.\enskip}
\def \Im{\mathop{\rm Im}\nolimits}
\def \Re{\mathop{\rm Re}\nolimits}
\def \mod{\mathop{\rm mod}\nolimits}
\newcount \eqnum
\def \chq#1{\global \advance \eqnum1 \xdef#1{(\the \eqnum)}#1}
\def\cheqno{\eqno\chq}

\def \q#1 {\bigskip \noindent #1.\enskip}

\def \pts#1 {\smallskip\rightline {(#1 pts)}}

\newcommand{\MM}{L}
\newcommand{\A}{\mbox{{\LARGE{\it a}}}}
\newcommand{\sA}{\mbox{{\large{\it a}}}}
\newcommand{\D}{\Delta}

\title{Hyperbolic Geometry of Kuramoto Oscillator Networks%
\thanks{Submitted on 1/14/2017.
		\funding{This work was supported by NSF Grant DMS 1413020}}}
\author{Bolun Chen\footnotemark[2]
\and  Jan R.~Engelbrecht%
\thanks{Department of Physics, Boston College, Chestnut Hill MA
	(\email{jan@bc.edu})}
\and Renato Mirollo%
\thanks{Department of Mathematics, Boston College, Chestnut Hill MA}
	}
\begin{document}
\maketitle

\begin{abstract}
Kuramoto oscillator networks have the special property that their trajectories
are constrained to lie on the (at most) 3D orbits of the M\"obius group acting
on the state space $T^N$ (the $N$-fold torus).  This result has been used to
explain the existence of the $N-3$ constants of motion discovered by Watanabe
and Strogatz for Kuramoto oscillator networks.  In this work we investigate
geometric consequences of this M\"obius group action.  The dynamics of Kuramoto
phase models can be further reduced to 2D reduced group orbits, which have a
natural geometry equivalent to the unit disk $\D$ with the hyperbolic metric.
We show that in this metric the original Kuramoto phase model (with order
parameter $Z_1$ equal to the centroid of the oscillator configuration of points
on the unit circle) is a gradient flow and the model with order parameter
$iZ_1$ (corresponding to cosine phase coupling) is a completely integrable
Hamiltonian flow.  We give necessary and sufficient conditions for general
Kuramoto phase models to be gradient or Hamiltonian flows in this metric.  This
allows us to identify several new infinite families of hyperbolic gradient or
Hamiltonian Kuramoto oscillator networks which therefore have simple dynamics
with respect to this geometry.  We prove that for the $Z_1$ model, a generic 2D
reduced group orbit has a unique fixed point corresponding to the hyperbolic
barycenter of the oscillator configuration, and therefore the dynamics are
equivalent on different generic reduced group orbits.  This is not always the
case for more general hyperbolic gradient or Hamiltonian flows; the reduced
group orbits  may have multiple fixed points, which also may bifurcate as the
reduced group orbits vary.  
\end{abstract} 



\section{Introduction}

Coupled oscillator networks 
are used to model a wide variety of interesting collective phenomena
in science and nature.
Examples include
synchronization of cardiac pacemaker cells and firefly populations\cite{winfree1967biological,mirollo1990synchronization},
dynamics of Josephson junction arrays\cite{nichols1992ubiquitous,strogatz1993splay}, 
electro-chemical oscillations\cite{zhabotinsky1991history},
synchronization of people walking \cite{strogatz2005theoretical}
etc.
This paper concerns a highly idealized class of oscillator networks, governed by equations of the form
$$
 \dot \theta_j = A + B \cos \theta_j + C \sin \theta_j, \quad j = 1, \dots, N. \cheqno\Ksys
$$ 
Here $\theta_j$ is an angular variable (i.e.~an element of ${\mathbb R}  \mod 2\pi \mathbb Z$) and the coefficients $A, B, C$ are smooth functions of $(\theta_1, \dots, \theta_N)$.  The state space for this system is the $N$-fold torus $T^N = (S^1)^N$.  We like to call an individual oscillator governed by an equation of the form above a {\sl Kuramoto oscillator}, and so we will refer to the oscillator networks defined above as {Kuramoto oscillator networks}.  If the functions $A, B, C$ are symmetric, i.e.~invariant under all permutations of the variables $\theta_j$, then we would call this a symmetric network of Kuramoto oscillators.  But we emphasize that we do not assume symmetry throughout this paper; the functions $A, B, C$ may depend differently on the $\theta_j$, or even not depend at all on some of the $\theta_j$.

Kuramoto oscillator networks  arise as  models of Josephson junction series arrays,
and also as the result of averaging more complex dynamical 
systems\cite{swift1992averaging}.  Beginning with the original work of Kuramoto over forty years ago \cite{kuramoto1975self,sakaguchi1986soluble}, Kuramoto networks have been a very fertile research subject in applied dynamics  
(reference \cite{pikovsky2015dynamics} is a nice survey of much of this work through 2015).
As these networks were extensively studied, researchers began to realize that Kuramoto oscillator systems exhibited dynamical properties that would be considered atypical in more general oscillator networks.  In particular, it became clear that the long-term dynamics often were neither asymptotically stable nor unstable; instead, a remarkable neutral stability for steady states was often observed.

A major step in understanding this neutral stability was achieved by Watanabe and Strogatz  in their 1994 paper ``Constants of motion for superconducting Josephson
arrays'' \cite{watanabe1994constants} which we will henceforth refer to as WS.  This seminal work is now considered one of the most important papers on the dynamics of Kuramoto networks.  In an algebraic tour-de-force, WS constructs $N-3$ independent functions which are conserved quantities for a system of the form \Ksys.  Therefore the dynamical orbit of any initial point in $T^N$ is constrained to lie on an at most 3-dimensional submanifold defined by setting these $N-3$ functions equal to constants. The WS theory was subsequently generalized to
non-identical oscillator networks\cite{pikovsky2008partially,vlasov2015star},
networks with external periodic forcing\cite{pikovsky2009self} and
noisy oscillators\cite{braun2012global}.
Furthermore, it is shown in \cite{pikovsky2011dynamics} that in the continuum limit $N\to\infty$,
the WS theory can be linked to the famous Ott-Antonsen ansatz\cite{ott2008low,ott2009long}, which is a
low-dimensional dynamical reduction technique that made possible the complete analytic solution to
numerous variations of the classic continuum limit Kuramoto model,
as in 
\cite{abrams2008solvable,martens2009exact,laing2009chimera,marvel2009invariant}.
More recently, the WS formalism has been extended perturbatively to weakly
inhomogeneous populations of Kuramoto oscillators\cite{vlasov2016dynamics}.

This reduction to 3D dynamics essentially explained the observed neutral stability of some steady states for Kuramoto networks.  For example, in the case of symmetric coefficient functions $A, B, C$ we are interested in {\sl splay} orbits, which are periodic dynamical orbits in which the angular variables $\theta_j$ all evolve according to the same periodic function, but with equally spaced time shifts.  Before WS, splay orbits were observed in Josephson junction networks and observed numerically to be neutrally stable in $N-2$ independent directions \cite{nichols1992ubiquitous,strogatz1993splay}.  In light of WS, this makes perfect sense; the splay orbits live inside 3D submanifolds defined by the WS constants of motion; perturbing in $N-3$ independent directions given by changing the WS constants results in an orbit constrained to lie on a different 3D submanifold, which cannot relax back to the original splay orbit.  (The remaining neutral direction to bring the count up to $N-2$ is the direction along the orbit itself.)

The next step forward was the realization that the WS constants have an intrinsic group-theoretic interpretation, and in fact it is this group action which is fundamental to the special dynamical properties of Kuramoto networks. The 3D group $G$ consisting of M\"obius transformations that preserve the unit disc acts naturally on $T^N$.  
In 2009 \cite{marvel2009identical} Mirollo, Marvel and Strogatz observed that the dynamical orbits of \Ksys\ are constrained to lie on the group orbits for this action. Therefore the dynamical system reduces to a family of 3D systems on the group orbits.   The WS constants can be interpreted as cross-ratios of points on the unit circle, which are preserved by M\"obius transformations.  We see this M\"obius group invariance as the intrinsic reason for the reduction to 3D dynamics, and think of the WS constants more as a consequence derived from the group action.
The M\"obius invariance 
also 
leads to a complete 
classification of attractors for 
Kuramoto networks
\cite{engelbrecht2014classification}.

But there is much more in WS than the constants of motion.  WS goes on to derive the evolution equations for the reduced dynamics on the 3D orbits, which we will present below in a more transparent M\"obius formulation.  Next, WS analyzes a special case of \Ksys\ obtained by Swift et.~al.~\cite{swift1992averaging} via averaging more general Josephson junction array systems; namely, the system given by
$$
 \dot \theta_j = \omega + {1 \over N} \sum_{k = 1}^N  \cos (\theta_k - \theta_j - \delta), \quad j = 1, \dots, N \cheqno\Kphase
$$ 
where $\omega$ and $\delta$ are constants.  This system has an additional invariance  given by $\theta_j \mapsto \theta_j + c$ for any $c \in \mathbb R$; if $(\theta_1(t), \dots \theta_N(t))$ is a solution then so is  $(\theta_1(t)+c, \dots \theta_N(t)+c)$.  So we can identify points $(\theta_1, \dots \theta_N)$ and $(\theta_1+c, \dots \theta_N+c)$ to obtain a reduced state space $\widetilde{T^N}$ which is topologically an $(N-1)$-dimensional torus, and the system dynamics will lie on the at most 2 dimensional reduced group orbits.

WS constructs a function $\cal H$ on each reduced group orbit with the property that $\dot {\cal H} = R^2 \sin \delta$, where $R$ is the magnitude of the centroid of the points $e^{i\theta_j}$.  This function $\cal H$ is a Lyapunov function for the flow unless $ \sin \delta = 0$. In the case $\sin\delta=0$,  $\cal H$ is an additional conserved quantity and therefore the system is completely integrable.  The dynamics on the reduced group orbits can be easily understood in terms of the function $\cal H$; in particular one can show that fixed points correspond to critical points of $\cal H$.  Closed orbits are ruled out unless $\sin \delta = 0$.  WS establishes that $\cal H$ has at least one critical point on the reduced group orbit of any $p \in T^N$ unless $p$ has a {\sl majority cluster} of at least $N/2$ identical $\theta_j$.  It is conjectured in WS that this critical point is unique; we will prove below that this is indeed correct.

One of the main results of this paper is to show that the system \Kphase\ with $\delta = \pm \pi/2$ is in fact a gradient flow on the reduced group orbits, with respect to a natural metric which is equivalent to the hyperbolic metric on the unit disc, and $\cal H$ can be derived as the potential function for this gradient flow.  In fact, this derivation is equivalent to the standard multivariable calculus problem of determining that a vector field is a gradient, and then integrating to find the potential function. Moreover, the flow for general $\delta$ is just a rotation of the gradient case with respect to this metric; in particular, the $\pm \pi/2$ rotation of the system corresponding to $\sin \delta = 0$ is Hamiltonian with respect to this metric. But most importantly, the system \Kphase\ is only one example of a Kuramoto network with this gradient/Hamiltonian structure.  We will exhibit a simple criterion for a Kuramoto network to have this property, and give several examples of Kuramoto networks for which the gradient/Hamiltonian dynamics hold.  We leave as an open problem the complete classification of Kuramoto networks with this gradient/Hamiltonian structure.

The organization of this paper is as follows: we begin by deriving the explicit equations for the dynamics on the 3D M\"obius orbits, then turn to the special case of systems with the additional invariance $\theta_j \mapsto \theta_j+c$, for which an additional reduction to 2D orbits holds.  We show that these 2D orbits are naturally equivalent to the unit disc with the standard hyperbolic metric, and derive a criterion for when the flow on the 2D reduced orbits is gradient with respect to this metric.  The special case \Kphase\ studied in WS has a particularly nice geometric interpretation in this metric, which we explain.  We prove the uniqueness of fixed points for \Kphase, as conjectured in WS, and then give several other examples of systems satisfying the gradient/Hamiltonian criterion.  We conclude with some discussion of directions for further research on these systems.

\section{Reduction To 3D System}

We begin by deriving the M\"obius form of the evolution equations 3.6 in WS, which give the dynamics on the group orbits. It is desirable to express the system \Ksys\ in complex form, with $z_j = e^{i \theta_j}$.  Let ${\A} = -C + i B$; $\A$ a complex-valued function on $T^N$ which plays the role of an order parameter for the system.  Then using $\dot z_j = i z_j \dot \theta_j$ we obtain
$$
\dot z_j =i A z_j +i z_j {\rm Im} ( {\A} \overline z_j ) =  i A z_j + {1 \over 2} \left( {\A} - \overline {\A} z_j^2 \right), \quad j = 1, \dots, N. \cheqno\KsysC
$$
As an example, the WS system \Kphase\ has ${\A} = e^{i(\pi/2 - \delta)} Z_1$, where $Z_1$ is the first moment of the point $(z_1, \dots, z_N)$ given by
$$
Z_1 = {1 \over N} \sum_{j = 1}^N z_j. 
$$
Henceforth we will refer to \Kphase\ as the $Z_1$ model.

Let $G$ be the 3D group $G$ of M\"obius transformations preserving the unit disc.   An element $M \in G$ can be expressed uniquely in the form
$$
Mz = \zeta {z - w \over 1 - \overline w z}, \quad z \in \mathbb C, \cheqno\Mob
$$
where the parameters $w$ and $\zeta$ satisfy $|w| < 1$ and $|\zeta| = 1$.  Therefore $G$ is topologically the product of the unit disc $\Delta$ and unit circle $S^1$.  Note that in this parameterization $w$ is the pre-image of $0$:  $w = M^{-1}(0)$ or equivalently $Mw= 0$.    When $\zeta = 1$, we denote the above M\"obius transformation by $M_w$.  If $M \in G$ and $p = (\beta_1, \dots, \beta_N) \in T^N$ then
$$
Mp = (M\beta_1, \dots, M\beta_N)
$$
defines the group action of $G$ on $T^N$.  The group orbits are the sets $Gp = \{Mp \ | \ M \in G\}$.

Now fix a base point $p = (\beta_1, \dots, \beta_N)  \in T^N$.  As shown in \cite{marvel2009identical}, any trajectory for \KsysC\ with initial condition in the group orbit $Gp$ can be expressed in the form $M(t) p$ for some $M(t) \in G$; we will explicitly derive this result below.  Let $M(t) \in G$ be any smooth $1$-parameter family with parameters $w =w(t)$ and $\zeta =\zeta (t)$, and let $z_j = M(t)\beta_j$ be the coordinates of $M(t)p$.
We differentiate $z_j = M(t) \beta_j$ directly to obtain
$$
\begin{aligned}
\dot z_j&= \dot \zeta {\beta_j  -w \over 1 - \overline{w} \beta_j } + \zeta \left ({ -\dot w \over 1 - \overline{w}  \beta_j  }\right) - \zeta \left( { \beta_j - w \over (1 - \overline w \beta_j)^2} \right)  (-\dot{\overline w} \beta_j) \cr
&= \dot \zeta \overline \zeta z_j -\left( { \dot w \over \beta_j - w} \right) z_j + \left( {\overline \zeta \dot{\overline w} \beta_j \over \beta_j - w} \right)z_j^2 \cr
&= \dot \zeta \overline \zeta z_j -\left( { \dot w \over \beta_j - w} \right) z_j 
+\overline \zeta \dot{\overline w}  \left( 1 + {w \over \beta_j - w} \right) z_j^2.
\end{aligned}     \cheqno\dotzj
$$
Inverting the equation for $z_j = M(t)\beta_j$ gives
$$
\beta_j = { z_j + \zeta w \over \zeta +  \overline w z_j}, \quad \quad {1 \over \beta_j - w} = {\zeta +\overline w z_j \over z_j(1-|w|^2)}
$$
which we substitute in \dotzj\ to obtain 
$$
\dot z_j=-{ \dot w \zeta \over 1 - |w|^2} + \left( \dot \zeta \overline \zeta+ {\dot{\overline w} w - \dot w \overline w \over 1 - |w|^2} \right ) z_j + {\dot{\overline w} \overline \zeta \over 1 - |w|^2} z_j^2.
$$
Comparing this to \KsysC, we see that if we set
$$
\begin{aligned}
\dot w &=  -{1 \over 2} ( 1 - |w|^2 )\overline{\zeta} {\A} \cr 
\dot \zeta &= iA \zeta - {1 \over 2} \left( \overline w{\A} - w \overline {\A} \zeta ^2  \right),
\end{aligned}
\cheqno\Eqwzeta
$$
with $A$ and $\A$ evaluated at the point $M(t)p = (z_1(t), \dots, z_N(t))$, then $M(t)p$ satisfies \KsysC.  Equation \Eqwzeta\ defines a dynamical system on the M\"obius group $G$ (which is topologically $\Delta \times S^1$).  If the base point $p$ has at least three distinct coordinates $\beta_j$, then any point in the group orbit $Gp$ has a {\sl unique} expression $Mp$ for some $M \in G$; this is because a M\"obius map is uniquely determined by the images of three distinct points.  So the system dynamics on the group orbit $Gp$ are equivalent to the dynamics on the group $G$ given by \Eqwzeta.

The factor $1-|w|^2$ in the $w$ equation is the first hint that this flow has connections to hyperbolic geometry, since $1-|w|^2$ is the denominator in the hyperbolic metric on the unit disc $|w| < 1$.  We also observe that if we express $M(t)p = \zeta M_wp$, then the $w$-equation takes the form
$$
\dot w = -{1 \over 2} (1-|w|^2) \overline \zeta {\A} (\zeta M_wp). \cheqno\Eqw
$$

\section{Change Of Base Point}

We explained above how to introduce coordinates $w \in \Delta$ and $\zeta \in S^1$ on any $G$-orbit $Gp$, provided that the point $p = (\beta_j) \in T^N$ has at least three distinct $\beta_j$, which we will assume from here on.  In this section we consider the effect of changing the base point $p$ to a different point $p' = \MM p$ in $Gp$.  Let $w', \zeta'$ be the coordinates associated to the base point $p'$.  If $q = Mp$ is any point in $Gp$, then for this point $q$, $w = M^{-1} (0)$.  Similarly, if $q = M' p'$, then for this point $q$, $w' = (M')^{-1}(0)$.  Now
$$
q = Mp = M' \MM p \Longrightarrow M = M'\MM \Longrightarrow M' = M \MM^{-1};
$$
therefore
$$
w' = (M')^{-1} (0) = \MM M^{-1} (0) = \MM w.
$$
This shows that the coordinates $w$ and $w'$ are related via the M\"obius transformation $\MM$; this observation will be crucial later in our discussion of hyperbolic geometry.

There is no similar simple relation between the coordinates $\zeta$ and $\zeta'$; since we will not need the precise relation in the sequel, we omit this derivation.  Note that we could have replaced $\zeta$ by the coordinate $\eta = M^{-1}(1) \in S^1$; then the change-of-coordinate rule is the same as for the $w$ coordinates: $\eta'   = \MM \eta$.  We chose to use $\zeta$ instead of $\eta$ to keep the form of the M\"obius transformation associated to $w$ and $\zeta$ in \Mob\ as simple as possible.  

\section{Kuramoto Phase Models}

It is tempting to cancel the $\zeta$ and $\overline \zeta$ in \Eqw, thus uncoupling the $w$ equation from $\zeta$; this is legitimate if $\A$ satisfies the invariance relation ${\A}(\zeta p) = \zeta {\A} (p)$.  This invariance relation holds if the system \Ksys\ is a {\sl Kuramoto phase model}, which we define to be a Kuramoto model with the additional property that 
if $\theta_j (t) $ is any solution, then so is $ \theta_j(t) + c$ for any constant $c$.
It is easy to see that this condition holds if and only if the defining functions (in complex form) satisfy the homogeneity relations  $A(\zeta p) = A(p)$ and ${\A}(\zeta p) = \zeta {\A}(p)$ for all $p \in T^N$ and all $\zeta$ with $|\zeta| = 1$.  The WS system \Kphase\ is an example: here $A = \omega$ 
and ${\A} = e^{i\alpha} Z_1$, 
which clearly satisfy the homogeneity conditions   
($\alpha=\pi/2 - \delta$ in terms of the parameter $\delta$ used in WS).
More generally, define the $n$th moment of the point $(z_1, \dots, z_N)$ for any $n \in \mathbb Z$ as
$$
Z_n = {1 \over N} \sum_{j = 1}^N z_j^n.
$$
Then we can construct a symmetric Kuramoto phase model by taking $\A$ to be 
any linear combination of terms
$$
Z_{n_1} \cdots Z_{n_r} \quad {\rm with } \quad n_1+ \dots +n_r = 1.
$$

For a Kuramoto phase model, the equation for $w$  uncouples from $\zeta$ and has the particularly simple form
$$
\dot w =  -{1 \over 2} ( 1 - |w|^2 ) {\A} ( M_w p).  \cheqno\weq
$$
The dynamics for a phase model can be further reduced to 2D, by identifying points under rotation; in other words, we identify $p$ and $\zeta p$ for any $\zeta \in S^1$.  The full state space for this reduced model is an $(N-1)$-dimensional torus; the group orbits $Gp$ under this identification give us reduced group orbits $\widetilde{Gp}$, which are invariant under the reduced dynamics.  For a base point $p$ with at least three distinct coordinates, its reduced $G$-orbit can be parametrized by $w \in \Delta$, and equation \weq\ gives the dynamics on the reduced orbit.  Note that the function $A$ is irrelevant to the dynamics for the reduced model.  We also remark  that fixed points in the reduced system correspond to either fixed points or uniformly rotating solutions (i.e.~constant phases) in the original $N$-dimensional system.

The Poincar\'e model for hyperbolic geometry on the unit disc $\Delta$  has metric
$$
ds = {2 | dw| \over 1-|w|^2}.
$$
This metric is conformal with the Euclidean metric (i.e.~angle measures agree), has constant negative curvature $-1$ and its geodesics are lines or arcs of circles which meet the boundary in $90^o$ angles. Since the reduced $G$-orbits are in one-to-one correspondence with $\Delta$ via the coordinate $w$, we can transfer this metric to the reduced $G$-orbits.  This metric on the reduced $G$-orbits is natural in the sense that it is {\sl independent} of the choice of base point.  This is because the orientation-preserving isometries for the Poincar\'e geometry are precisely the M\"obius transformations in our group $G$.  If we change base points, then the relation between the $w$ and $w'$ coordinates is given by a M\"obius transformation, which preserves the hyperbolic metric.  

\section{Gradient Condition}

Since the metric on the reduced $G$-orbits is intrinsically defined, it is natural to explore connections between the dynamics of these reduced systems and the associated geometry given by the metric.  In particular, one of the simplest things that could happen is that the dynamical system is a gradient system with respect to this metric.  So we ask, when is \weq\ a gradient flow for the hyperbolic metric?  Recall that if $w = u+iv \in \Delta$, then for any smooth function $h$ on $\Delta$ we define the complex partial derivatives
$$
{\partial h \over \partial w} = {1 \over 2} \left ( {\partial h \over \partial u} - i  {\partial h \over \partial v}\right), \quad {\partial h \over \partial \overline w} = {1 \over 2} \left ( {\partial h \over \partial u} + i  {\partial h \over \partial v}\right).
$$
Then the Euclidean gradient of a real function $h$ in complex form is given by
$$
\nabla_{euc} h = 2 {\partial h \over \partial \overline w}.
$$
In general, the gradient of a real function $h$ with respect to a conformal metric $\phi \, ds$, where $ds$ is the ordinary Euclidean metric on ${\mathbb R}^N$, is given by $\phi^{-2} \nabla_{euc} h$, where $\nabla_{euc} h$ is the ordinary Euclidean gradient of $h$.
So the hyperbolic gradient of $h$ is given by 
$$
\nabla_{hyp} h = {1 \over 4}(1-|w|^2)^2 \nabla_{euc} h = {1 \over 2}(1- |w|^2)^2{\partial h \over \partial \overline w}.
$$

Now consider a dynamical system on $\Delta$ in complex form 
$$
\dot w = f(w) = U + i V
$$
with $U, V$ real.  Then
$$
{\partial f \over \partial w} = {1\over 2} \left[ {\partial U \over \partial u} + {\partial V \over \partial v} + i \left ( {\partial V \over \partial u} - {\partial U \over \partial v}\right)\right],
$$
so the Euclidean gradient condition in complex form is just 
$$
\Im {\partial f \over \partial w} = 0.
$$
Similarly, the hyperbolic gradient condition for $f$ is 
$$
\Im \left ( {\partial \over \partial w} \left[ (1-|w|^2)^{-2} f(w) \right] \right) = 0. \cheqno\hypgrad
$$

Suppose $\dot w = f(w)$ satisfies the hyperbolic gradient condition on $\Delta$; then one can construct a real function $h$ on $\Delta$, unique up to a constant, such that $f = \nabla_{hyp} h$. Then along trajectories,
 $$
 \dot h (w)= || \nabla_{hyp}h(w) ||_{hyp}^2 = (1 - |w|^2)^2 \left| {\partial h \over \partial \overline w} \right|^2.
 $$
Next, suppose we rotate the vector field $f$ by some 
fixed  $\zeta = e^{i \alpha} \in S^1$; 
in other words, we consider the flow $\dot w = \zeta f(w)$.  Then along trajectories we have
$$
\dot h(w) = \langle \nabla_{hyp}h(w), \zeta f(w) \rangle _{hyp} = \cos \alpha\; (1 - |w|^2)^2 \left| {\partial h \over \partial \overline w} \right|^2.
$$
Thus we see that provided $\cos \alpha \ne 0$, the function $h$ is strictly increasing or decreasing along trajectories (except for fixed points of the flow).  In the case $\alpha = \pm \pi/2$ the function $h$ is a conserved quantity, and in fact the flow is Hamiltonian with respect to the hyperbolic metric, with Hamiltonian function $h$. 
The system is completely integrable in the Hamiltonian case, with trajectories defined by the level  curves of $h$.

For the reduced system \weq,
which has 
$$
f(w)=-\frac12 (1-w\bar w)\A(M_w p)
$$  
the hyperbolic gradient condition is
$$
\Im
\left(
{\partial\over \partial w}
[(1-w\bar w)^{-1}\A(M_w p)]
\right)
=0.
$$
We have
$$
{\partial\over \partial w}
[(1-w\bar w)^{-1}\A(M_w p)]
=(1-w\bar w)^{-2}\bar w\A(M_w p)
+
(1-w\bar w)^{-1}\sum_{j=1}^N{\partial\A\over \partial z_j}{\partial z_j\over \partial w}
.
\cheqno\ten
$$
Here
the base point $p=(\beta_1,\beta_2,\ldots,\beta_N)$
and
$$
z_j=M_w\beta_j={\beta_j-w\over 1-\bar w\beta_j},
$$
so
$$
{\partial z_j\over \partial w}=-{1\over 1-\bar w\beta_j}.
\cheqno\dzjdw
$$
For a phase model the order parameter $\A$ satisfies the homogeneity condition 
$$
\A(\zeta z_1,\zeta z_2, \ldots, \zeta z_N)
=\zeta
\A(z_1,z_2, \ldots, z_N);
$$
differentiating with resepct to $\zeta$ gives the identity
$$
\sum_{j=1}^N z_j{\partial\A\over\partial z_j}=\A.
\cheqno\homogeneous\
$$
Substituting
\dzjdw\
and
\homogeneous\
into
\ten\ gives
$$
\begin{aligned}
{\partial\over \partial w}
[(1-w\bar w)^{-1}\A(M_w p)]
&= 
(1-w\bar w)^{-2}
\sum_{j=1}^N
\left(
\bar w z_j
-{1-w\bar w\over 1-\bar w\beta_j}
\right)
{\partial\A\over \partial z_j}
\\
&=
(1-w\bar w)^{-2}
\sum_{j=1}^N
\left(
\bar w \cdot
{\beta_j-w\over 1-\bar w\beta_j}
-{1-w\bar w\over 1-\bar w\beta_j}
\right)
{\partial\A\over \partial z_j}
\\
&=
-(1-w\bar w)^{-2}
\sum_{j=1}^N
{\partial\A\over \partial z_j}
.
\end{aligned}
$$
Since $1-w\bar w$ is real,
we see that the hyperbolic gradient condition is 
$$
\Im D {\A} = 0 \cheqno\eqgrad
$$
everywhere on $T^N$,  
where
the differential operator $D$ on the torus $T^N$ with coordinates $z_j \in S^1$ is
$$
 D = {\partial \over \partial z_1} + \cdots + {\partial \over \partial z_n}.
$$
The flow for the system \weq\ is Hamiltonian for the hyperbolic metric if and only if
the flow with order paramater $i\A$ is gradient, so the hyperbolic Hamiltonian condition is
$$
\Re D {\A} = 0 
.
$$

The function $\A=Z_1$ from the WS system \Kphase\ with $\alpha=0$ ($\delta=\pi/2$)  
satisfies the hyperbolic gradient criterion:  $D {\A} = 1$, so $ \Im D {\A} = 0$.  
This special case of the original Kuramoto model \Kphase, 
with $\omega=0$, is also a gradient system on the full state space $T^N$ 
with respect to the standard Euclidean metric 
$ds^2=
d\theta_1^2
+\ldots +
d\theta_N^2
$;
its potential function (up to a constant) is
$
(N/2)|Z_1|^2
$.
However, in general
the hyperbolic gradient condition \eqgrad\ is not equivalent to the Euclidean gradient condition
on $T^N$.
For example, 
the system \Ksys\ with order parameter 
$
\A=|Z_1|^2Z_1
$
and $A=0$ is gradient with respect to the Euclidean metric on $T^N$, but this $\A$
does not satisfy the hyperbolic gradient condition \eqgrad.
Conversely, 
the system \Ksys\ with  
$
\A=Z_2\overline{Z_1}
$
and $A=0$,
where 
$
Z_2 = {1 \over N} \sum_{j = 1}^N z_j^2,
$
is not gradient with respect to the Euclidean metric on $T^N$, but does
satisfy the hyperbolic gradient condition \eqgrad.
We will present several additional examples 
of hyperbolic gradient systems
in Section 8.
 
\section{$\bf Z_1$ Phase Model}

The $Z_1$ phase model \Kphase\ studied in WS has
$
{\A} = e^{i\alpha}Z_1,
$
so the dynamics on the reduced orbits are given by
$$
\dot w =  -{1 \over 2} ( 1 - |w|^2 )e^{i\alpha} Z_1( M_w p).  \cheqno\sweq
$$

\begin{figure}[h]
	\centerline{
		\includegraphics[height=1.7in]{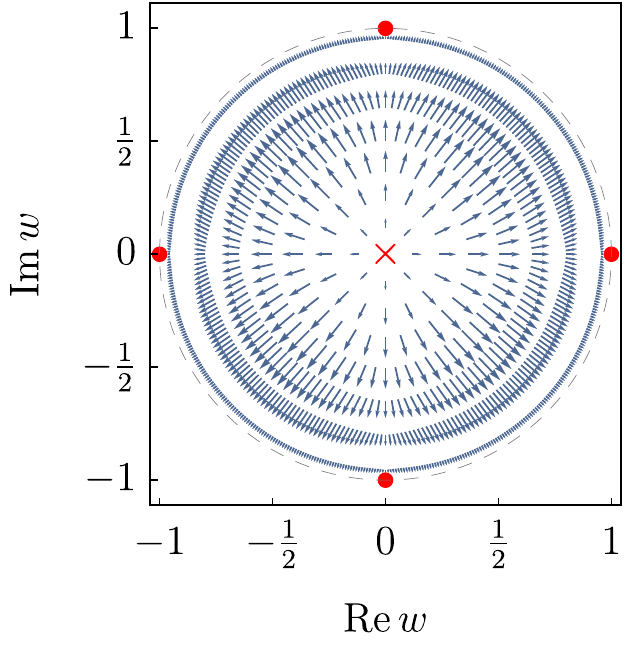}
		\includegraphics[height=1.7in]{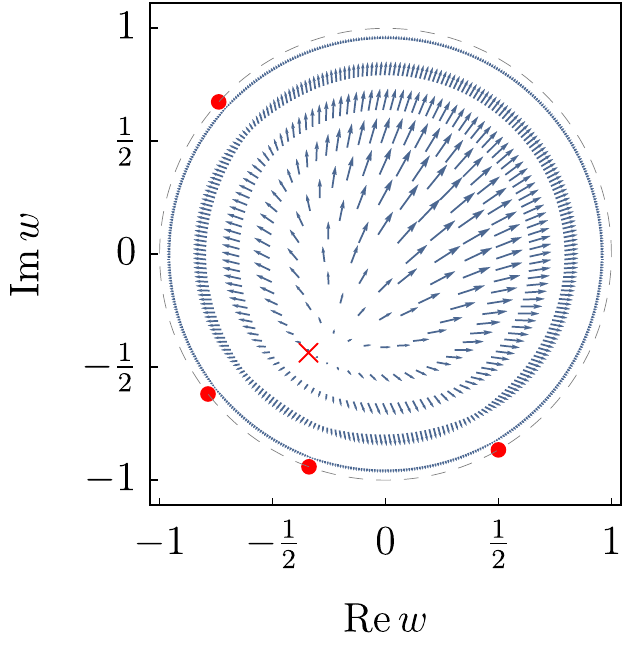}
		\includegraphics[height=1.7in]{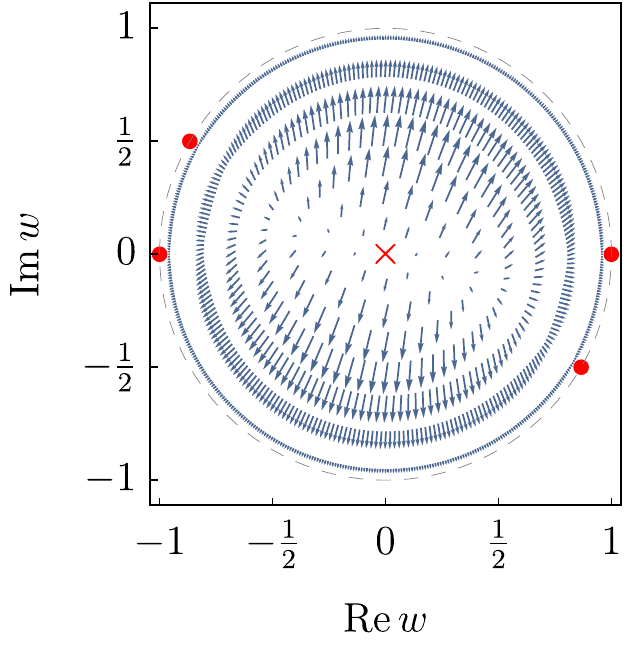}
	}
	\vspace{-11pt}
	\caption{\label{fig1}
		Vector fields $\dot{w}$ on $\D$  
		corresponding to flows on reduced $G$-orbits
		with base points 
		$p_A$, $p_B$ and $p_C$
		for the  phase model 
		$\sA=Z_1$ with $N=4$.
		The dots are the coordinates of the base point $p$; $\times$ is the hyperbolic barycenter.
	}
\end{figure}

It is illustrative to plot the
vector fields $\dot{w}$ on $\D$  
which correspond to the flows on  reduced $G$-orbits of the oscillator system
described by the  phase model $\A=Z_1$ with $N=4$.
Figure~\ref{fig1} shows the fields for the base points
$p_A=(1,i,-1,-i)$,
$p_B=M_wp_A$ with $w=0.5\;e^{i\pi/3}$,
and
$p_C=(1,\eta,-1,-\eta)$
with $\eta=e^{i5\pi/6}$.
Panels A) and B) 
are equivalent flows related by the M\"obius transformation $M_w$;
Panel A) is more symmetrical since its
base point  has barycenter at zero.  
Panel C represents the flow on
a different reduced group orbit 
with base point $p_C$ which can be thought of as
a deformation of $p_A$  fixing the  barycenter at zero. 
For the $Z_1$ model the flows on the reduced group orbits are topologically equivalent,
provided the base point $p$ has all distinct coordinates.

As shown above, \sweq\ is a hyperbolic gradient system when $\alpha=0$, and so has a potential function $\cal H$.  Comparing to \hypgrad, we see that we can construct $\cal H$ by solving
$$
\begin{aligned}
{\partial {\cal H} \over \partial \overline w} &= -(1-|w|^2)^{-1} Z_1(M_wp) 
\cr 
&= -{1 \over N} \cdot{1 \over 1 - w \overline w} \sum_{j=1}^N {\beta_j - w \over 1-\overline w \beta_j} \cr
&= {1 \over N} \sum_{j=1}^N\left( { w \over 1- w\overline w} -{\beta_j \over 1 - \beta_j \overline w} \right).
\end{aligned}
$$
Integrating with respect to $\overline w$, treating $w$ as a constant, determines $\cal H$ up to an arbitrary analytic function $g(w)$.  We obtain
$$
{\cal H} (w) =  {1 \over N} \sum_{j=1}^N \log \left( {1 - \beta_j \overline w \over 1 - w \overline w} \right) + g(w).
$$
Next, we want to  choose $g(w)$ to make $\cal H$ real, so we set
$$
\begin{aligned}
{\cal H} (w) &=  {1 \over N} \sum_{j=1}^N \left[\log \left( {1 - \beta_j \overline w \over 1 - w \overline w} \right) + \log (1 - \overline \beta_j w) \right]\cr
&= - {1 \over N} \sum_{j=1}^N \log \left( {1- |w|^2  \over (1 - \beta_j \overline w) (1 - \overline \beta_j  w)} \right) .
\end{aligned}
$$
Let $\rho_\beta(w)$ denote the Poisson kernel function with unit mass at $\beta \in S^1$:
$$
\rho_\beta(w) = { 1 - |w|^2 \over (1- \overline \beta w) ( 1- \beta \overline w)}  = { 1 - |w|^2 \over 1 - 2 {\rm Re} \overline \beta w + |w|^2} .
$$
Recall that $ {1 \over 2 \pi} \rho_z(w)$ is a density function on the circle $|w| = r < 1$, and these densities converge to the delta function at $\beta$ as $r \to 1$.  Then we see that the potential function $\cal H$ is the negative average of logs of Poisson densities:
$$
{\cal H} (w) =- {1 \over N} \sum_{j = 1}^N \log \rho_{\beta_j} (w),
$$
Using the notation of WS, 
the $Z_1$  model with 
${\A} = e^{\left ( \pi / 2 - \delta \right) i }Z_1 $
has
$$
\dot {\cal H} (w)= \cos \left ( {\pi \over 2} - \delta \right)(1-|w|^2)^2 \left| {\partial {\cal H}\over \partial \overline w} \right|^2 = \sin \delta |Z_1(M_wp)|^2,
\cheqno\Heq
$$
in agreement with WS.  

\begin{figure}[h]
	\centerline{
		\includegraphics[height=1.7in]{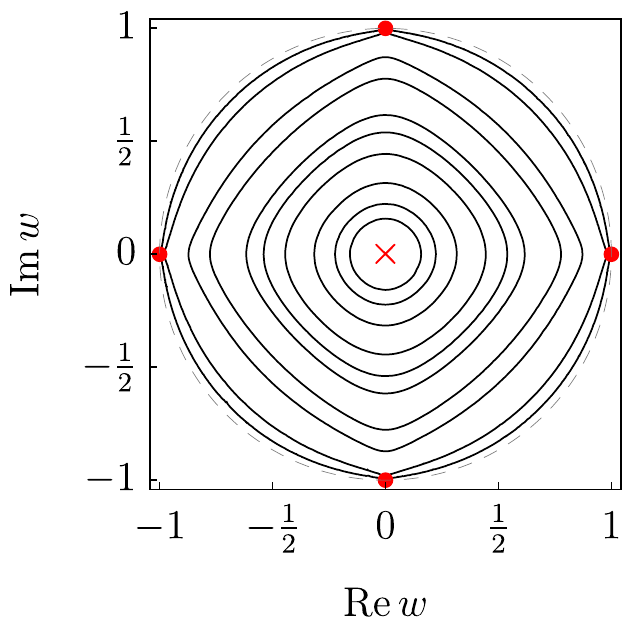}
		\includegraphics[height=1.7in]{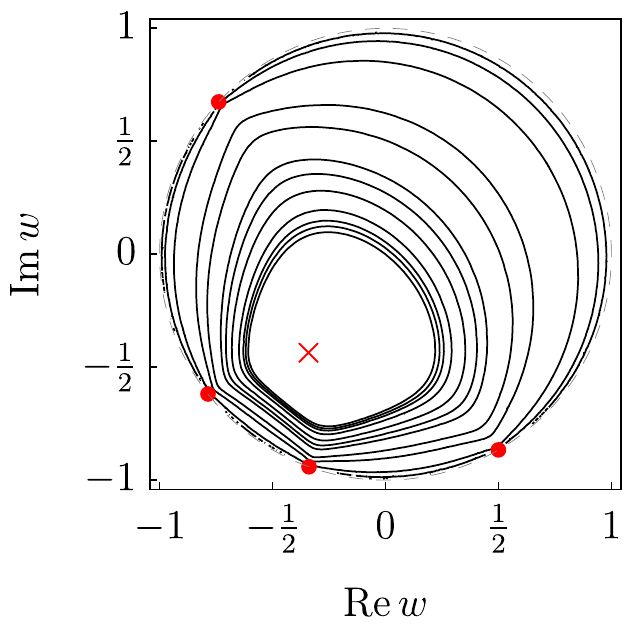}
		\includegraphics[height=1.7in]{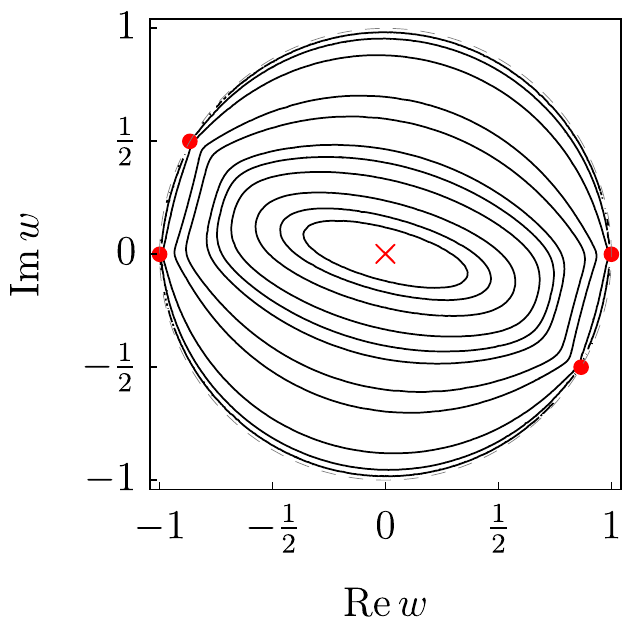}
	}
	\vspace{-11pt}
	\caption{\label{fig2}
		Level curves of the Hamiltonian $\cal H$
		for base points 
		$p_A$, $p_B$ and $p_C$
		as in Figure~\ref{fig1}.
		Vector fields in Figure~\ref{fig1} are the hyperbolic
		gradients of $\cal H$.
	}
\end{figure}

\begin{figure}[h]
	\centerline{
		\includegraphics[height=1.7in]{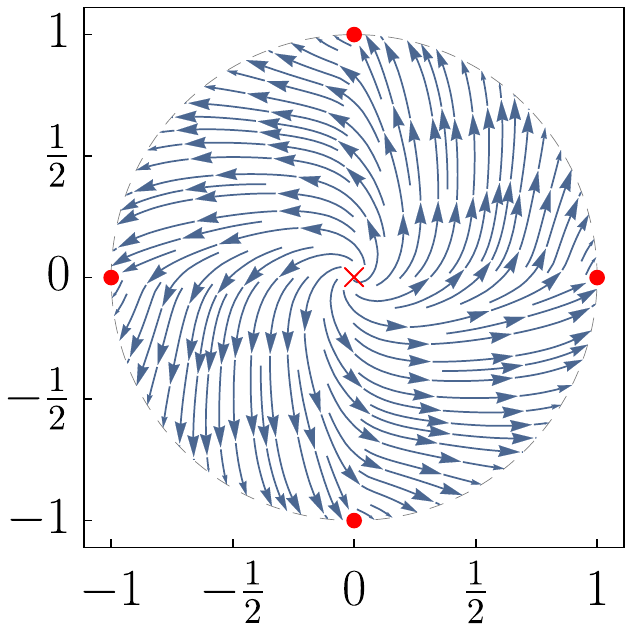}
		\includegraphics[height=1.7in]{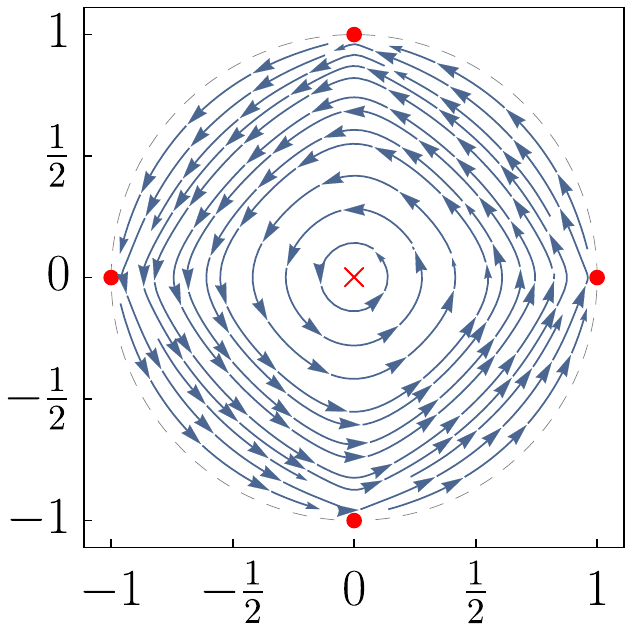}
	}
	\vspace{-11pt}
	\caption{\label{fig3}
                Vector fields $\dot{w}$ on $\D$
                corresponding to flows on $G$-orbits
                with basis point
                $p_A$,
		for the model 
		$\sA=e^{i\alpha}Z_1$ and $N=4$
                with rotation parameter $\alpha=\pi/4$ in Panel A
                and $\alpha=\pi/2$ in Panel B.
	}
\end{figure}

In Figure~\ref{fig2} we plot level curves of the Hamiltonian function ${\cal H}(w)$ 
for the three base points $p_A$, $p_B$ and $p_C$ used in Figure~\ref{fig1}. The flows in Figure~\ref{fig1} (with $\alpha=0$)
are the hyperbolic gradients of $\cal H$.
For $\alpha\ne0$ the vector fields on $\D$ are rotated by the angle
$\alpha$  from the hyperbolic gradient:
$\dot w=e^{i\alpha}\nabla_{hyp}{\cal H}$. 
For $\alpha=\pm\pi/2$
the flow is along level curves of ${\cal H}$.
Figure~\ref{fig3}  depicts vector fields on the reduced $G$-orbit for $N=4$ 
with base point $p_A$.
The rotation parameter $\alpha=\pi/4$ in Panel A
yields outwardly spiraling dynamics with  increasing ${\cal H}$.
Panel B shows the completely integrable Hamiltonian case $\alpha=\pi/2$ with
${\cal H}$ conserved.

\section{Geometric Interpretation of ${\cal H}$ for $\A=e^{i\alpha}Z_1$ model}

\begin{figure}
	\centerline{
		\includegraphics[height=1.7in]{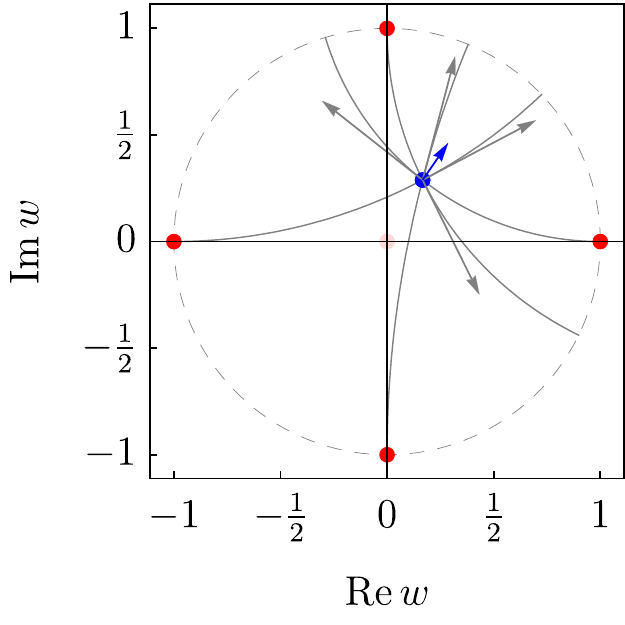}
		\includegraphics[height=1.7in]{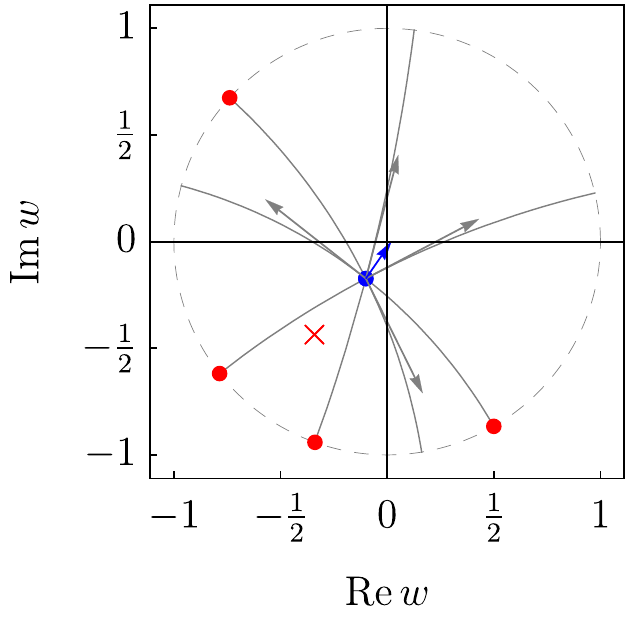}
		\includegraphics[height=1.7in]{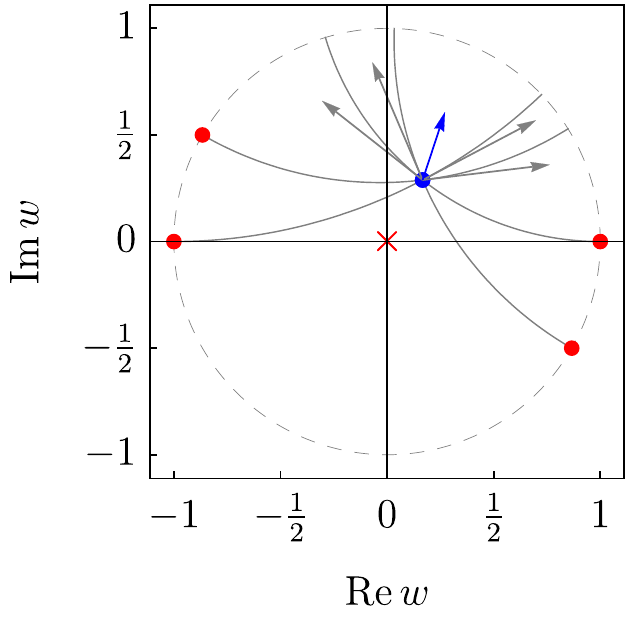}
	}
	\vspace{-11pt}
	\caption{\label{fig4}
                Geodesics connecting
                a point $w\in\D$
                to the four $\beta_j$'s of the base points $p_A$, $p_B$ and $p_C$.
                Grey arrows indicate each geodesic's unit direction $\xi_{\beta_j}(w)$, they sum to the blue arrow which is the direction $\dot w$
                for the phase model 
		$\sA=Z_1$.
	}
\end{figure}

 The flow on the disc $\Delta$ given by \sweq\ and the function $\cal H$ have beautiful interpretations in terms of the hyperbolic geometry on the disc,   explained in the 1986 paper ``Conformally Natural Extension Of Homeomorphisms Of The Circle'' by Douady and Earle \cite{douady1986conformally}.   (Note that Douady and Earle omit the factor $2$ in the definition of the hyperbolic metric, so their metric has curvature $-4$.)  Fix any point $\beta$ on the boundary $S^1$; then for each $w \in \Delta$ there is a unique geodesic that connects $w$ to $\beta$.  Therefore for each $w \in \Delta$ there is a unique unit vector (in the hyperbolic metric)  $\xi_\beta(w)$ which gives the direction of the geodesic connecting $w$ to $\beta$; the corresponding geodesic flow is given by
$$
\dot w  = \xi_\beta(w) = {1 \over 2} (1-|w|^2) {\beta-w \over 1-\overline w \beta} = {1 \over 2}(1-|w|^2) M_w \beta.
$$
For example, suppose $\beta = 1$ and $w=x \in (-1,1)$; then the flow reduces to 
$$
\dot x = {1 \over 2} \left( 1 - x^2\right),
$$
which is exactly the flow on $(-1,1)$ towards $1$ with unit speed in the hyperbolic metric.
The vector field $\xi_\beta$ is the hyperbolic gradient of the real function $h_\beta$ given by
$$
h_\beta(w) =  \log \rho_\beta (w) = \log \left( {1 - |w|^2 \over |\beta - w|^2} \right).
$$
So we see that the $Z_1$ model \sweq\ with $\alpha=0$ (which has ${\A} = Z_1$)  is just the average of these geodesic flows towards the points $z_j$, reversed in time, and is the gradient flow for 
$$
{\cal H}(w) = -{1 \over N} \sum_{J = 1}^N h_{\beta_j}(w).
$$
This is illustrated in Figure~\ref{fig4}, where we plot the 
four geodesics connecting a point $w$ to the four $\beta_j$'s for
the base points $p_A$, $p_B$ and $p_C$ in each panel respectively.
The unit geodesic directions $\xi_{\beta_j}$ at $w$ are shown as the grey vectors
which sum to the blue vector which in turn indicates the direction of the
flow $\dot w$.

The unique fixed point for the flow \sweq\ is the {\sl conformal barycenter} of the configuration $p =(\beta_1, \dots, \beta_N)$ on the unit circle.  This point is defined by the property that at this point the sum of the unit vectors pointing towards the $\beta_j$ is $0$.  Douady and Earle prove the existence and uniqueness of the conformal barycenter for a continuous probability distribution on the circle, and assert that their proof can be modified for the case of discrete masses, as long as there are no atoms with mass $\ge 1/2$.

There is also a nice interpretation, due to Thurston\cite{douady1986conformally}, of the functions $h_\beta$; roughly speaking, $-h_\beta(w)$ measures the distance from $w$ to $\beta$ relative to the distance from $0$ to $1$.  Of course both these distances are infinite in the hyperbolic metric, so more precisely, this means
$$
-h_\beta(w) = \lim_{r \to 1^-} \left( d(w, r\beta) - d(0, r) \right).
$$
Therefore ${\cal H}(w)$ measures (in this relative sense) the average distance from $w$ to the points $\beta_j$ on the boundary.  The conformal barycenter for the configuration $p= (\beta_j)$ is the unique point which minimizes $\cal H$.

We conclude this section with a proof  that the $Z_1$ model has a unique fixed point on each reduced group orbit $\widetilde{Gp}$, provided that $p$ does not have a {\sl majority cluster} of at least $N/2$ equal $\beta_j$.  (WS proves existence but not uniqueness).  Suppose $p$ is a fixed point for the reduced system, and $p$ has no majority cluster, so $p$ must have at least $3$ distinct $\beta_j$.  Construct the function $\cal H$ as above.  Existence and uniqueness are a consequence of the following lemmas:

\begin{lemma}  
	If $p$ has no majority cluster, then
$$
\lim_{|w| \to 1} {\cal H}(w) = \infty.
$$
\end{lemma}  

\begin{lemma}  
All fixed points of \sweq\ for $\cos\alpha < 0$ 
are attracting.
\end{lemma}  

Assume these lemmas hold and let $w \in \Delta$ be any point. Consider the forward limit set $\Omega(w)$ under the flow \sweq\ with $\cos\alpha<0$.  Then $\cal H$ is decreasing (or constant) along the trajectory of $w$, so Lemma 1 implies that the forward limit set $\Omega(w)$ must be a compact subset of $\Delta$.  Then $\cal H$ takes a minimum value over $\Omega(w)$ at some point $w^\star$, and we see from \Heq\ that $w^\star$ must be a fixed point for the flow.  By Lemma 2 all fixed points are attracting, so we must have $\Omega(w) = \{ w^\star \}$.  This proves the existence of fixed points, and also that each $w \in \Delta$ is in the basin of attraction of some fixed point.  If there were multiple fixed points, we would have a partition of $\Delta$ into disjoint non-empty open basins of attraction, which is impossible.  This proves uniqueness.

\begin{proof}[Proof of Lemma 1]  The assertion is equivalent to
$$
\lim_{|w| \to 1} e^{-N{\cal H}(w)} = 0.
$$
So it suffices to prove that
$$
\lim_{n \to \infty} e^{-N{\cal H}(w_n)} = 0
$$
for any sequence $w_n \in \Delta$ with  $\lim w_n  \in S^1$.
Observe that
$$
e^{-N{\cal H}(w_n)} = \prod_{j = 1}^N \rho_{\beta_j} (w_n) = \prod_{j = 1}^N { 1 - |w_n|^2 \over |1 - \overline \beta_j w_n|^2}.
$$
If $\lim w_n  \ne \beta_j$ for all $j$, then as $n \to \infty$ the denominators in all the factors are bounded below by some $c > 0$, so the conclusion is clear.  Otherwise suppose $\lim w_n = \beta_j$, and $\beta_j$ occurs with multiplicity $l$ in $p$.  Observe that for $0 < r < 1$, 
$$
\max_{|w| = r} \rho_{\beta_j}(w) = {1+r \over 1-r} < {2 \over 1-r}.
$$
Therefore up to constants the $j$ term is dominated by $(1-|w_n|)^{-l}$ and all the other terms together are dominated by $(1-|w_n|)^{N - l}$; hence as long as $l < N/2$ we have $
\lim_{n \to \infty} e^{-N{\cal H}(w_n)} = 0$.
\end{proof}

\begin{proof}[Proof of Lemma 2]  Suppose the reduced phase model has a fixed point $p=(\beta_1,\beta_2,\ldots,\beta_N)$.  We choose $p$ as our base point and consider the system \sweq. To first order in $w$, 
$$
M_w \beta_j = (\beta_j - w)(1 + \overline w \beta_j) = \beta_j - w + \beta_j^2 \overline w,  
$$
so	
the linearization of \sweq\  at $w = 0$ is 
$$
\dot w = -{1 \over 2}e^{i\alpha}  \left ( Z_2 \overline w  - w\right),
$$
where $Z_2$ is the second moment of $p$.  Let $Z_2 = a + i b$; then in real coordinates $w = u + iv$ this 2D linear system has matrix 
$$
L = -{1 \over 2}\begin{pmatrix} \cos \alpha & - \sin\alpha \cr \sin \alpha & \cos \alpha\end{pmatrix} \begin{pmatrix} a - 1 & b \cr b & -a -1\end{pmatrix}.
$$
Observe that $\rm{tr} \, L = \cos \alpha $ and $\det L = {1 \over 4} (1 - |Z_2|^2) > 0$, so the fixed point at $w = 0$ is attracting when $\cos \alpha < 0$.  
\end{proof}

In terms of the $N=4$ examples used for illustrative purposes, if the point $w\in\D$ in each panel of Figure~\ref{fig4} at which the tangent vectors $\xi_{\beta_j}(w)$
are evaluated is taken to be at $w=Z_1(p)$, pairs of $\beta_j$ fall on the same geodesic but with opposite flow directions, so the $\xi_{\beta_j}(w)$ come in canceling pairs and
the barycenter $w$ is a fixed point.

\section{New Families of Hyperbolic Gradient Phase Models}

We have shown that the widely studied $Z_1$ model is a hyperbolic gradient 
phase model, which clarifies some of its special properties discovered
in WS.  It is not unique.
As stated earlier 
one can construct Kuramoto phase models by taking $\A$ to be
any linear combination of terms
$
Z_{n_1} \cdots Z_{n_r} \quad {\rm with } \quad n_1+ \dots +n_r = 1
$,
but these models generally do not 
satisfy the hyperbolic gradient condition \eqgrad.
We have, however,
identified infinite families of such gradient phase models
that can be written as combinations of double, triple and quadruple products of moments, namely
$$
\begin{aligned}
{\cal D}_n &=
        Z_n Z_{1-n}, \\
{\cal T}_n &=
Z_{1+2n}
Z_{-n}^2
-
Z_{1-2n}
Z_{n}^2
+Z_{1+n}Z_{n}Z_{-2n}
-Z_{1-n}Z_{-n}Z_{2n}
,\\
\mathcal{Q}_{n} & =
\left(
Z_{1+n}Z_{-n}
-
Z_{1-n}
Z_{n}
\right)
|Z_{n}|^2
\end{aligned}
$$
where $n \in \mathbb Z$ is arbitrary. 
The $Z_1$ model is then a special case of the first (double product) family with $n=1$.
All of these are easy to check, using these facts: the differential operator $D$ is a derivation, $D Z_n = n Z_{n-1}$ and $\overline {Z_n} = Z_{-n}$.
For example, ${\A} = Z_n Z_{1-n}$ has
$$
D(Z_n Z_{1-n}) = nZ_{n-1}  Z_{1-n} + (1-n) Z_n Z_{-n} = n | Z_{n-1}|^2 + (1-n) |Z_n|^2
$$
which is real, so $\Im D{ \A} = 0$.

\begin{figure}
\centerline{
	\includegraphics[height=1.70in]{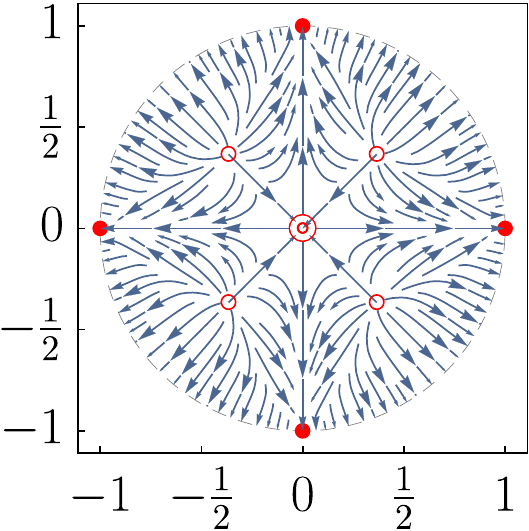}
    \includegraphics[height=1.70in]{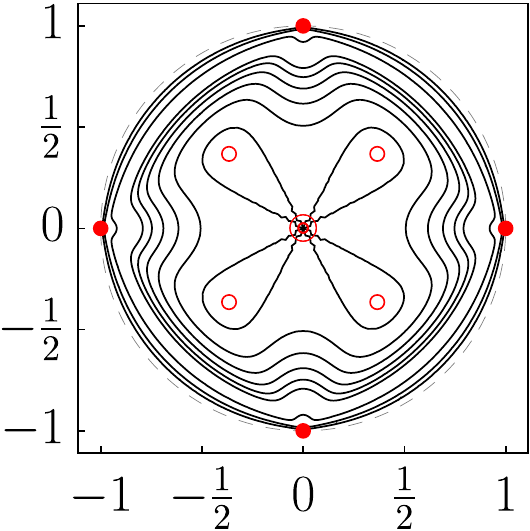}
	\includegraphics[height=1.70in]{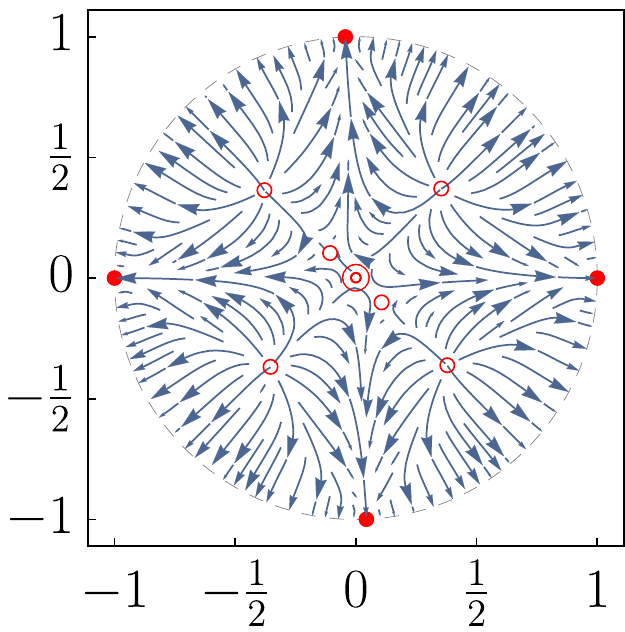}
}
\vspace{-11pt}
\caption{\label{fig5}
		Vector fields $\dot{w}$ on $\D$  
		corresponding to flows on reduced $G$-orbits
		with base points 
		$p_A$ and $p_C'$
		in Panels A and C.
		Panel B shows
		level curves of the potential $\cal H$
		whose gradient is the flow in panel A.
		These results are all for the phase model 
		$\sA=Z_2Z_{-1}$ with $N=4$.
		The dots are the coordinates of the base points;  circles indicate fixed points.
	}
\end{figure}

We mention
some properties of the simplest extension of the $Z_1$ model, with
$
\A=e^{i\alpha}Z_2Z_{-1}.
$
The function $Z_2Z_{-1}$ can have multiple
zeros on reduced $G$-orbits.
For instance, in Panel A of Figure~\ref{fig5} we plot the flow for the gradient case
$\alpha=0$ corresponding to splay $G$-orbit for the $p_A$ base point for $N=4$.
Notice there are now five fixed points inside the disk; one  at the barycenter $w=0$ and
four at $\pm w^{*}$ and $\pm \overline{w^*}$ with $w^{*}=\sqrt{2-\sqrt{3}}\;e^{i\pi/4}$.  
The fixed point at $w=0$ is non-hyperbolic with index $-3$; the other four fixed points 
are hyperbolic with index $+1$.  
We have also calculated the potential $\cal H$ for 
$\A=Z_2Z_{-1}$, and  plot level sets of $\cal H$ for the base point $p_A$ in panel B.
In Panel C we plot the gradient flow corresponding to the
different reduced $G$-orbit with base point  
$p_C'=(1,i\eta',-1,-i\eta')$
where $\eta'=e^{i\pi/72}$.
The fixed point
at  $w=0$ (with index $-3$) from Panel A has now bifurcated into 3 hyperbolic fixed points.
So we see that for this model, fixed point bifurcations can occur as we vary the base point $p$,
in contrast to the case of the $Z_1$ model.

\section{Discussion}

In this paper we have presented a new framework for studying the dynamics of Kuramoto phase models.  For a system with $N$ oscillators, the phase space for these systems reduces to the torus $T^{N-1}$, and the dynamical orbits lie in the reduced M\"obius group orbits, which generically can be identified with the unit disc $\Delta$. The reduced M\"obius orbits have a natural hyperbolic metric, so there is an interesting subset of Kuramoto phase models which are gradient systems with respect to this metric.  An example is the $Z_1$ model studied in WS.  We showed that most of the special dynamical properties of the $Z_1$ model reported in WS are consequences of this hyperbolic gradient structure.  We presented a simple criterion for Kuramoto phase models to have this gradient property, and gave several families of such models. We leave as an open problem the complete classification of these hyperbolic gradient systems.

The dynamics of Kuramoto phase models with the gradient property, and more generally their rotations with respect to the intrinsic hyperbolic metric, can be analyzed fairly easily in terms of the potential function associated to the flow; we hope to present some examples of this for some of the gradient systems we gave above in future work.  For a complete dynamical picture, it is necessary to include the boundaries of the reduced $G$-orbits, which generically consist of $N$ copies of the circle $S^1$, corresponding to states with all but one of the oscillators in sync, 
which we call $(N-1,1)$ states.
These $N$ circles all meet in a single point corresponding to the completely in-sync state.  These boundary circles are invariant under the dynamics, and typically contain saddle points which determine separatrices for the dynamics in the reduced $G$-orbits.
For example, the $Z_1$ model with $\sin \delta > 0$ has a single repelling fixed point (the conformal barycenter) in each reduced $G$-orbit; there are $N$ heteroclinic saddle connections joining the barycenter to $N$ saddles, one on each boundary component.  All other trajectories converge to the in-sync state on the boundary.  The dynamics are reversed for $\sin \delta < 0$, and Hamiltonian for $\sin \delta = 0$.

This dynamical portrait is discussed in WS, where it is stated 
``On each invariant subspace, the flow is either toward the in-phase state
(if $\sin \delta > 0$), toward the incoherent manifold ($\sin \delta < 0$), or neither ($\sin \delta = 0$).''
(The ``incoherent manifold'' is the codimension 2 set of all conformal barycenters.)
This description is almost correct, but misses the codimension one
manifolds connecting the barycenters to the $(N-1,1)$ boundary saddles.
In any case, the dynamics on each reduced group orbit is qualitatively the same; there are no bifurcations as one moves through the reduced group orbits.  
This is definitely not the case for more complicated gradient phase models; 
interesting bifurcations can occur as we vary the orbits.  
For example, 
as we saw above for the $Z_2Z_{-1}$ model,
if the base point $p$ is a highly symmetric configuration like the $N$th roots of unity, than the fixed point at $w = 0$ on the reduced $G$-orbit can be non-hyperbolic, and bifurcate to multiple fixed points as we vary the base point $p$.  
We plan to address this and other issues related to the dynamics of these gradient systems in a future work.

We thank Steve Strogatz for suggesting that we revisit some of the questions raised in WS, 
Martin Bridgeman for pointing out reference \cite{douady1986conformally}, and both of them
for many helpful discussions while this work was in progress.

\bibliographystyle{siamplain}
\bibliography{refs}

\end{document}